\documentclass[aps,superscriptaddress,twocolumn,showpacs]{revtex4}
\usepackage{amsmath}
\usepackage{graphicx}
\begin{document}
\title{Novel Attractive Force Between Ions in Quantum Plasmas}
\author{P. K. Shukla}
\affiliation{International Centre for Advanced Studies in Physical Sciences \&  Institute for Theoretical Physics, 
Faculty of Physics \& Astronomie, Ruhr-University Bochum, D-44780 Bochum, Germany; Department of Mechanical 
and Aerospace Engineering \& Center for Energy Research, University of California San Diego, La Jolla, CA 92093, U. S. A.}
\email{profshukla@yahoo.de}
\author{B. Eliasson}
\affiliation{International Centre for Advances Studies in Physical Sciences \& Institute for Theoretical
Physics, Faculty of Physics \&  Astronomy, Ruhr-University Bochum, D-44780 Bochum, Germany}
\email{beliass@yahoo.se}
\received{12 December 2011}
\revised{19 June 2012}
\begin{abstract}
We report new attractive force between ions that are shielded by degenerate electrons 
in quantum plasmas. Specifically, we show that the electric potential around an isolated 
ion has a hard core negative part that resembles the Lennard-Jones (LJ)-type potential. Physically, 
the new electric potential is attributed to the consideration of the quantum statistical pressure, 
the quantum Bohm potential, as well as the electron exchange and electron correlations due to 
electron-1/2 spin within the framework of the quantum hydrodynamical description of quantum plasmas. 
The shape of the attractive potential is determined by the ratio between the Bohr radius and the 
Wigner-Seitz radius of degenerate electrons. The existence of the hard core negative potential
will be responsible for the attraction of ions forming lattices and atoms/molecules, as well as 
for critical points and phase transitions in quantum plasmas at nanoscales.    
 \end{abstract}
\pacs{71.10.Ca, 63.10.+a, 67.10.Hk}

\maketitle

A study of the potential distribution around a test charged object is of fundamental importance in many 
physical systems (e.g. dilute charge stabilized colloidal suspensions such as latex spheres in water,
micelles and microemulsions, condensed matters with strongly correlated electrons and holes, strongly 
coupled dusty and quantum plasmas in laboratory and astrophysical environments, etc.), since its knowledge 
predicts how a cloud of opposite polarity charges will shield a test charge particle over a certain 
radius, which is known as the Debye-H\"uckel radius in the context of electrolytes and plasmas and the 
Thomas-Fermi/Yukawa radius in the context of condensed matters. The traditional repulsive screened Coulomb  potential assumes 
the form $\phi (r) =(Q/r)\exp(-r/\lambda)$, where $Q$ is the test charge and $\lambda$ the  shielding radius of the sphere. In a 
classical electron-ion plasma \cite{r1,r2}, an isolated ion is shielded by  non-degenerate Boltzmann distributed electrons, 
and hence one \cite{r3} replaces $Q$ by  $Z_i e$ and $\lambda$  by the electron Debye radius 
$\lambda_{De}= (k_B T_e/4\pi n_0e^2)^{1/2}$, where $Z_i$ is the ion charge state, 
$e$ the magnitude of the electron charge, $k_B$ the Boltzmann constant, $T_e$ the electron temperature, 
and $n_0$ the unperturbed electron number density. For a slowly moving test charge in collisionless \cite{Montgomery68} 
and collisional \cite{Stenflo73,Yu73} plasmas, there appear additional far-field potentials decreasing as inverse cube 
and inverse square of the distance between the test charge and the observer. In a collisionless dusty plasma \cite{r4,r5,r5a} 
with Boltzmann distributed electrons and ions, a micron-sized negatively charged isolated dust is shielded by both 
positive ions and electrons. Here $Q$ equals $Z_d e$ and $\lambda$ is replaced by the effective dusty plasma Debye 
radius $\lambda_{D} = \lambda_{De}\lambda_{Di}/\sqrt{\lambda_{De}^2 + \lambda_{Di}^2}$, where $Z_d$ is the number 
of electrons residing on a dust grain, $\lambda_{Di} =(k_B T_i/4\pi n_{i0}e^2)^{1/2}$ the ion Debye radius, 
$T_i$ the ion temperature, $n_{i0} =n_0 + Z_d n_{d0}$ the ion number density, and $n_{d0}$ the dust number density. In dusty plasmas 
with $T_e \gg T_i$, we have $\lambda_D \approx \lambda_{Di}$. Furthermore, in dense Thomas-Fermi plasmas with 
an impurity ion (with the charge state $Z_*$) is shielded by non-relativistic degenerate electrons, so that 
$Q =Z_* e$ and $\lambda$ is replaced by the Thomas-Fermi radius \cite{r6} $\lambda_F = ({\cal E}_F/4\pi n_{0}e^2)^{1/2}$, 
where the electron Thomas-Fermi energy is denoted by ${\cal E}_F =(\hbar^2/2m_* k_B)(3\pi^2)^{2/3}n_0^{2/3}$, 
$\hbar$ is the Planck constant divided by $2\pi$, and $m_*$ the effective mass of the electrons (for example, 
for semiconductor quantum wells, we typically have $m_* =0.067 m_e$, where $m_e$ is the rest mass of the electrons).  

In the past, tremendous progress has been made in carrying out systematic theoretical and numerical studies of phase 
diagrams \cite{r7,r8,r9,r10,r11,r12,r12a} for colloidal systems, dusty plasmas, and strongly interacting matter by supposing 
that like-charged particles repel each other due to the Debye-H\"uckel/Yukawa repulsive force. However, besides 
the repulsive interaction, there are also attractive forces \cite{r4,r5a} between two like charged particles due 
to the overlapping Debye spheres \cite{r13} and due to the polarization of charged particulates by the sheath 
electric field \cite{r14,r15}. Henceforth, both short range repulsive and long-range attractive potentials play 
a decisive role in the theory and experiments of phase transitions in colloidal and dusty plasmas. 

In this Letter, we present a new attractive force between ions that are shielded by the degenerate electrons in 
strongly coupled quantum plasmas that are ubiquitous in a variety of physical environments (e.g. the cores of 
Jupiter and white dwarf stars \cite{r16,r17}, warm dense matters \cite{r17a}), in compressed plasmas produced by
intense laser beams \cite{r17b}, as well as in the processing devices for modern high-technology (e.g. semiconductors 
\cite{r18}, thin films and nano-metallic structures \cite{r19}, etc.) Specifically, we shall use here the generalized quantum 
hydrodynamical (G-QHD) equations \cite{r19} for non-relativistic degenerate electron fluids supplemented by 
Poisson's equation. The G-QHD model includes the quantum statistical pressure and the 
Bohm potential \cite{Gardner96,Manfredi01,Tsintsadze09}, as well as the electron exchange and electron correlation effects.
We demonstrate here that the electric potential around an isolated ion in quantum plasmas has a new distribution, 
the profile of which in special cases resembles the Lennard-Jones-type potential. It emerges that the newly 
found electric potential, arising from the static electron dielectric constant that includes the combined effects 
of the density perturbations associated with the quantum statistical pressure and the quantum force \cite{r20,r20b,r21} 
involving the overlapping of the electron wave function due to the Heisenberg uncertainty and Pauli's exclusion 
principles, as well as the electron exchange and electron correlation effects \cite{r22} due to electron-$1/2$ spin, 
embodies a short-range negative hard core electric potential. The latter will be responsible for Coulomb ion crystallization 
and oscillations of ion lattices under the new force associated with our exponential oscillating-screened Coulomb 
potential in  strongly coupled quantum plasmas. 

Let us consider a quantum plasma in the presence of non-relativistic degenerate electron fluids and mildly coupled ions that are
immobile and form the neutralizing background. In our quantum plasma, the electron and ion coupling parameters are $\Gamma_e
=e^2/a_e k_B T_F$ and $\Gamma_i =Z_i^2 e^2/a_i k_B T_i$, respectively, where $a_e  \sim a_i =(3/4\pi n_0)^{1/3}$ is the average 
interparticle distance, and $T_F =(\hbar^2/2m_* k_B)(3\pi^2 n_0)^{2/3}$ the Fermi electron temperature. It turns out that 
$\Gamma_i/\Gamma_e =Z_i^2 T_F/T_i \gg 1$, since in quantum plasmas we usually have $T_F > T_i$. The dynamics of degenerate 
electron fluids is governed by the continuity equation
\begin{equation}
  \frac{\partial n}{\partial t} + \nabla \cdot (n {\bf u}) =0,
\end{equation}
the momentum equation \cite{r19}
\begin{equation}
  m_* \bigg(\frac{\partial {\bf u}}{\partial t} + {\bf u}\cdot\nabla {\bf u}\bigg) 
= e \nabla \phi - n^{-1} \nabla P +\nabla V_{xc} + \nabla V_B,
\end{equation}
and Poisson's equation
\begin{equation}
  \nabla^2 \phi = \frac{4\pi e}{\epsilon} (n -n_0) - 4 \pi Q  \delta({\bf r}),
\end{equation}
where $n$ is the electron number density, ${\bf u}$ the electron fluid velocity, $\phi$ the electric potential,
and $\epsilon$ the relative dielectric permeability of the material (for example, for semiconductor quantum wells
we have $\epsilon = 13$). We have denoted the quantum statistical pressure 
$P=(n_0 m_* v_*^2 / 5) (n/n_0)^{5/3}$, where $v_*=\hbar (3\pi^2)^{1/3}/m_* r_0$ is the electron Fermi speed and 
$r_0 =n_0^{-1/3}$ represents the Wigner-Seitz radius, and the sum of the electron exchange and 
electron correlations potential  is \cite{r22} 
$V_{xc} =0.985(e^2/\epsilon)n^{1/3}\left[1+ (0.034 /a_B n^{1/3}){\rm ln}(1+18.37 a_B n^{1/3})\right]$, 
where $a_B =\epsilon \hbar^2/m_* e^2$ represents the effective Bohr radius. The quantum Bohm potential is \cite{r20,r20b,r21} 
$V_B = (\hbar^2/2m_*)(1/\sqrt{n}) \nabla^2 \sqrt{n}$. We have thus retained the desired quantum forces that act on degenerate 
electrons in our non-relativistic quantum plasma. The quantum hydrodynamic equations (1)-(3) are valid \cite{r19,r21a,r21b} if the 
plasmonic energy density $\hbar \omega_{pe}$ is smaller (or comparable) than (with) the Fermi electron kinetic energy $k_B T_F$, 
where $\omega_{pe}=(4 \pi n_0 e^2/\epsilon m_*)^{1/2}$ is the electron plasma frequency, and the electron-ion collision 
relaxation time is greater than the electron plasma period.  

Letting $n = n_0 + n_1$, where $n_1 \ll n_0$, we linearize the resultant equations to obtain the electron density 
perturbation $n_1$ that can be inserted into Eq. (3). The Fourier transformation in 
space leads to  the electric potential around an isolated ion. In the linear approximation, we have  \cite{Montgomery68,Else10}  
\begin{equation}
  \phi({\bf r})=\frac{Q}{2\pi^2} \int \frac{\exp(i {\bf k}\cdot {\bf r})}{k^2 D} d^3 k,
  \label{phi}
\end{equation}
where ${\bf r}$ denotes the position relative to the instantaneous position of the point test charge, and the dielectric constant 
for a dense quantum plasma with quasi-stationary density perturbations is given by
\begin{equation}
  D=1+\frac{\omega_{pe}^2}{k^2(v_{*}^2/3+v_{ex}^2)+\hbar^2k^4/4 m_*^2}.
\end{equation}
Here  we have denoted $v_{ex}=(0.328e^2/m_*\epsilon r_0)^{1/2}[1+0.62/(1+18.36a_B n_0^{1/3})]^{1/2}$.

The inverse dielectric constant can be written as
\begin{equation}
  \frac{1}{D}=\frac{(k^2/k_s^2) + \alpha k^4/k_s^4}{1+(k^2/k_s^2)+\alpha k^4/k_s^4},
  \label{D}
\end{equation}
where $k_s=\omega_{pe}/\sqrt{v_{*}^2/3+v_{ex}^2}$ is the inverse Thomas-Fermi screening length, 
and $\alpha={\hbar^2 \omega_{pe}^2}/{4 m_*^2 (v_{*}^2/3+v_{ex}^2)^2}$ measures the importance of the quantum recoil effect. 
If $m_*=m_e$, $\epsilon=1$ then $\alpha$ depends only on $r_0/a_B$ with $a_B=\hbar^2/m_e e^2$. 
Inserting Eq. (\ref{D}) into Eq. (\ref{phi}), the latter can be written as
\begin{equation}
  \phi({\bf r})
=\frac{Q}{4\pi^2}\int \bigg[ \frac{(1+b)}{k^2+k_{+}^2} 
+ \frac{(1-b)}{k^2+k_{-}^2} \bigg] \exp(i {\bf k}\cdot {\bf r}) d^3 k,
\label{pot}
\end{equation}
where $b =1/\sqrt{1-4\alpha}$, and $k_{\pm}^2={k_s^2}[1\mp\sqrt{1-4\alpha}]/2\alpha$.
Here, we use $\sqrt{1-4\alpha}=i\sqrt{4\alpha-1}$ for $\alpha > 1/4$.
The integral in Eq. (\ref{pot}) can be evaluated using the general formula
\begin{equation}
  \int \frac{\exp(i {\bf k}\cdot {\bf r})}{k^2+k_\pm^2}d^3 k= 2 \pi^2 \frac{\exp(-k_{\pm} r)}{r},
\end{equation}
where the branches of $k_\pm$ must be chosen with positive real parts so that the boundary condition
$\phi\rightarrow 0$ at $r\rightarrow \infty$ is fulfilled. 

\begin{figure}[htb]
\centering
\includegraphics[width=8.5cm]{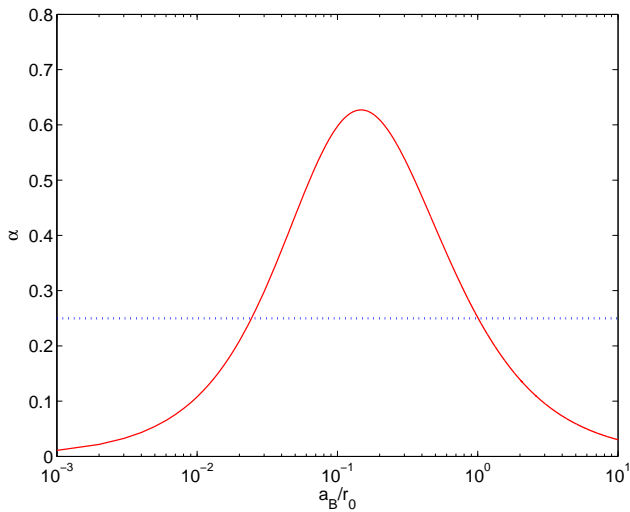}
\caption{The value of $\alpha$ as a function of $a_B/r_0$. The critical value $\alpha=1/4$ is indicated
with a dotted line.}
\end{figure}

First, for $\alpha> 1/4$, the solution of  the equation $k_\pm^2= k_s^2 (1\mp i\sqrt{4\alpha-1})/ 2\alpha$ 
yields $k_\pm=({k_s}/\sqrt{4\alpha})[(\sqrt{4\alpha}+1)^{1/2} \mp i (\sqrt{4\alpha}-1)^{1/2}] \equiv k_r\mp i k_i$,
and the electric potential
\begin{equation}
  \phi({\bf r})= \frac{Q}{r} \big[ \cos(k_i r) + b_\ast \sin(k_i r) \big] \exp(-k_r r),
  \label{pot1}
\end{equation}
where $b_\ast=1/\sqrt{4\alpha-1}$.
In the limit $\hbar^2 k^2 \gg 4 m_*^2 (v_*^2/3 + v_{ex}^2)$, we recover the exponential cosine-screened Coulomb potential \cite{r23}
\begin{equation}
  \phi({\bf r})=\frac{Q}{r} \cos(k_s \bar{r}) \exp(-k_s\bar{r}),
\end{equation}
where $\bar{r}=r/(4\alpha)^{1/4}$. Second, for $\alpha\rightarrow 1/4$, we have $k_+=k_-=\sqrt{2} k_s$, 
and 
\begin{equation}
  \phi({\bf r})= \frac{Q}{r} \bigg(1+\frac{k_s r}{\sqrt{2}}\bigg) \exp(-\sqrt{2}\,k_s r).
  \label{pot2}
\end{equation}
Third, for $\alpha < 1/4$, the expression $\sqrt{1-4\alpha}$ is real, and we
obtain $k_\pm=k_s (1\mp\sqrt{1-4\alpha})^{1/2}/\sqrt{2\alpha}$.
The resultant electric potential is 
\begin{equation}
  \phi({\bf r})=\frac{Q}{2 r} \big[ (1+b)\exp(-k_{+}r) + (1-b) \exp(- k_{-} r)
  \big].
  \label{pot3}
\end{equation}
We note that  in the limit $\alpha \rightarrow 0$, we recover from (\ref{pot3}) the modified Thomas-Fermi  screened Coulomb 
potential $\phi({\bf r})= (Q/r) \exp(-k_s r)$. Furthermore, the newly found electrical potential,  given by Eq. (12), is significantly 
different from that potential [e.g. Eq. (13) in Ref. \cite{Else10}, indicating that the electric potential is proportional to 
$r^{-3}\cos(2k_Fr)$, where $k_F =p_F/\hbar =|{\bf k}||2$ is the Fermi wave number  and $p_F$ the Fermi electron momentum] 
which involves the  Friedel oscillations \cite{Friedel58} arising from the Kohn  anomaly \cite{Kohn59} related with the discontinuous 
Fermi surface.

\begin{figure}[htb]
\centering
\includegraphics[width=8.5cm]{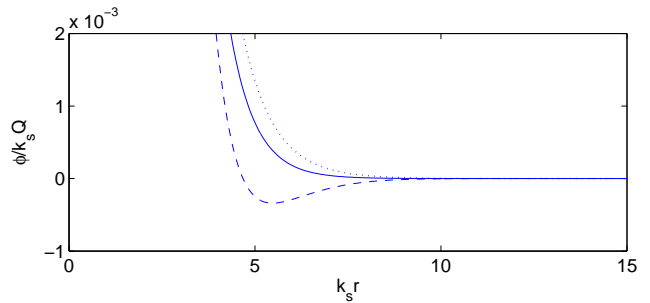}
\caption{The electric potential $\phi$ as a function of $r$ for $\alpha=0.625$ (dashed curve), 
$\alpha=1/4$ (solid curve) and $\alpha=0$ (dotted curve). The value 0.627 is the maximum possible value of $\alpha$ in our model,
obtained for $a_B/r_0\approx 0.15$.}
\end{figure}

\begin{figure}[htb]
\centering
\includegraphics[width=8.5cm]{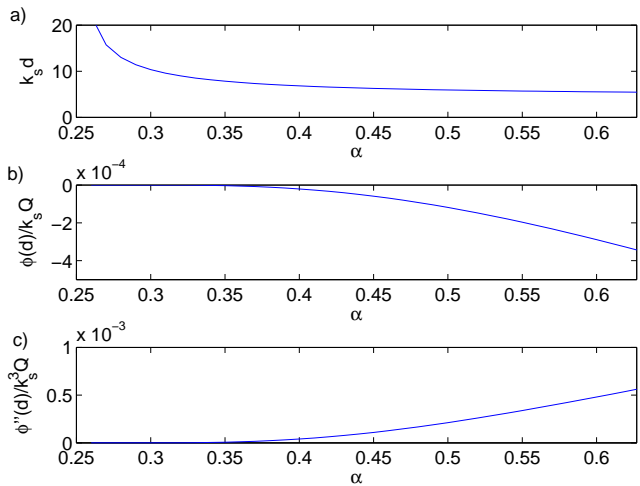}
\caption{a) The distance $r=d$ from the test ion charge where $d\phi/dr=0$ and the electric potential has 
its minimum, and b) the values of the potential $\phi$ and 
c) its second derivative $d^2\phi/dr^2$ at $r=d$.}
\end{figure}

Figure 1 shows the value of $\alpha$ as a function of $a_B/r_0$, with a maximum $\alpha$ at $a_B/r_0\approx 0.15$,
corresponding to a number density $n_0\approx 2\times 10^{22}\,\mathrm{cm}^{-3}$ (with $a_B=5.3\times10^{-9}\,\mathrm{cm}$), 
a few times below solid densities. The value of $\alpha$ is above the critical value $1/4$ only for a limited range of denities.
In Fig. 2, we display the profiles of the potential [given by Eq. (\ref{pot1}), (\ref{pot2}) and (\ref{pot3}) for
$\alpha>1/4$, $\alpha=1/4$ and $\alpha < 1/4$, respectively] for different values of $\alpha$. We clearly see the 
new short-range attractive electric potential that resembles the LJ-type potential for $\alpha > 1/4$, while
for smaller values of $\alpha$, the attractive potential vanishes. Figure 3(a)  depicts the distance $r=d$ from 
the test ion charge where $d\phi/dr=0$ and the electric potential has  its minimum. and Figs. 3(b) and 3(c) 
show the values of $\phi$ and $d^2\phi/dr^2$ at $r=d$.  The value of $(d^2\phi/dr^2)$ determines the 
oscillation frequency and dispersive properties of  the plasma as shown below. For $\alpha\leq 0.25$, 
the electric potential and its second derivative vanish,  and there is no attractive potential associated with 
the stationary test ion charge. Furthermore, we note that the  shielding of a moving test charge and bound 
states near a moving charge in a quantum plasma have been investigated  by Else {\it et al.} \cite{Else11} by 
using the Lindhard dielectric function \cite{Lindhard54} that ignores the electron exchange and electron correlation 
effects.  

The interaction potential energy between two dressed ions with charges $Q_i$ and $Q_j$ at the 
position ${\bf r}_i$ and ${\bf r}_j$ can be represented as $U_{i,j}({\bf R}_{ij})=Q_j \phi_i({\bf R}_{ij})$, where
$\phi_i$ is the potential around particle $i$, and ${\bf R}_{ij} ={\bf r}_i - {\bf r}_j$. For $\alpha > 1/4$, it reads, 
using Eq.(\ref{pot1}),
\begin{equation}
\begin{split}
&U_{i,j}({\bf R}_{ij}) =\frac{Q_i Q_j}{|{\bf R}_{ij}|}
\exp(-k_r |{\bf R}_{ij} |) 
\\
&\times\big[ \cos(k_i |{\bf R}_{ij}|) + b_\ast \sin(k_i |{\bf R}_{ij}|) \big].
\end{split}
  \label{potential}
\end{equation}
On account of the interaction potential energy, ions would suffer vertical oscillations around their 
equilibrium position.  The vertical vibrations of ions in a crystallized ion string in quantum plasmas 
are governed by
\begin{equation}
M \frac{d^2\delta z_j(t)}{d t^2} = - \sum_{i\neq j} \frac{\partial U_{ij}({\bf r}_i, {\bf r}_j)}{\partial z_j},
\label{newton}
\end{equation}
where $\delta z_j(t) [= z_j (t) -z_{j0}]$ is the vertical displacement of the $j$th ion from its equilibrium position
$z_{j0}$, and $M$ the mass of the ion. Assuming that $\delta z_j(t)$ is proportional to $\exp[-i (\omega t
- j k a)]$, where $\omega$ and $k$ are the frequency and wave number of the ion lattice oscillations, respectively,
and that $Q_i=Q_j=Q$, Eq. (\ref{newton}) for the nearest-neighbor ion interactions gives    
\begin{equation}
\omega^2 = \frac{4 Q^2}{M d^3} S  \exp(-k_r d) \sin^2 \left(\frac{k d}{2}\right),
\label{frequency1}
\end{equation}
where $S =  [2 (1+k_r d) + (k_r^2-k_i^2) d^2](\cos \xi + b_\ast \sin\xi) + 2 k_i d(1 + k_r d)
(\sin \xi - b_\ast \cos \xi)$.  $\xi =k_i d$, and $d$ is the separation between two consecutive ions. 
We note that Eq. (\ref{frequency1}) can also be obtained from the formula \cite{r24}
\begin{equation}
\omega^2 =\frac{4}{M} \left[\frac{d^2W (r)}{dr^2} \right]_{r=d} \sin^2 \left(\frac{k d}{2}\right),
\label{frequency2}
\end{equation}
where the inter-ion potential energy is represented as $W(r) =(Q^2/r)\exp(-k_r r)\left[\cos(k_i r) + b_\ast \sin(k_ir)\right]$ 
for $\alpha > 1/4$. Hence, the lattice wave frequency is proportional to $[{d^2\phi (r)}/{dr^2} ]_{r=d}^{1/2}$ 
[cf. Fig. 3(c)], and vanishes for $\alpha<0.25$.

Summing up, we have discovered a new attractive force between two ions that are shielded by degenerate  electrons 
in an unmagnetized quantum plasma. There are several consequences of our newly found short-range attractive 
force at quantum scales.  For example, due to the trapping of ions in the negative part of the exponential 
oscillating-screened Coulomb potential, there will arise ordered ion structures depending the electron density 
concentration,  which in fact controls the Wigner-Seitz radius $r_0$. The formation of ion clusters/ion atoms will 
emerge as  new features that are attributed to the new electric potential we have found here. Finally, under the action 
of the attractive force, we can have the formation of Coulombic ion lattices (Coulomb ion crystallization) and ion lattice 
vibrations, as well as the phenomena of phase separations at nanoscales in dense quantum plasmas (e.g. from solid to 
liquid-vapor phases) depending upon how one controls the ratio $r_0/a_B$. Thus, the ratio between the interaction 
energy between the two nearest-neighbor ions in the presence of the oscillating exponential Coulomb potential and 
the ion thermal energies, as well as the inter-particle spacing are the key parameters which will determine a  critical point 
that is required for phase transitions in quantum plasmas.  In conclusion, the present investigation, which has revealed the 
new physics of collective electron interactions at nanoscales, will open a new window for research  in one of the modern areas 
of physics dealing with strongly coupled degenerate electrons and non-degenerate mildly coupled ions in dense plasmas that 
share knowledge with cooperative phenomena (e.g. the formation of ion lattices) in condensed  matter physics and in astrophysics. 
Thus, the present investigation contributes to enhancing the existing knowledge of Wigner  crystallization in two-component 
Coulomb systems \cite{Bonitz05,Bonitz08} that do not account for an attractive force between  like charged particles.      

\acknowledgments
This work was supported by the Deutsche Forschungsgemeinschaft through the project SH21/3-2 of the Research Unit 1048.


\begin{thebibliography}{99}

\bibitem{r1} I. Langmuir, Phys. Rev. {\bf 33}, 954 (1929).
\bibitem{r2}W. B. Thompson and J. Hubbard, Rev. Mod. Phys. {\bf 32}, 714 (1960).
\bibitem{r3} R. O. Dendy, {\sl Plasma Dynamics} (Clarendon Press, Oxford, 1990).
\bibitem{Montgomery68} J. Neufeld and R. H. Ritchie, Phys. Rev. {\bf 98}, 1632 (1955);
G. Cooper, Phys. Fluids {\bf 12}, 2707 (1969); D. Montgomery {\it et al.}, 
Plasma Phys. {\bf 10}, 681 (1968).
\bibitem{Stenflo73} L. Stenflo and M. Y. Yu, Phys. Scr. {\bf 8}, 301 (1973).
\bibitem{Yu73} M. Y. Yu, R. Tegeback, and L. Stenflo, Z. Physik {\bf 264}, 341 (1973).
\bibitem{r4} P. K. Shukla and A. A. Mamun, {\sl Introduction to Dusty Plasmas} 
(Institute of Physics, Bristol, 2002).
\bibitem{r5} V. E. Fortov {\it et al.}, Phys. Usp. {\bf 47}, 447 (2004).
\bibitem{r5a} P. K. Shukla and B. Eliasson, Rev. Mod. Phys. {\bf 81}, 25 (2009). 
\bibitem{r6} E. M. Lifshitz and L. P. Pitaevskii, {\sl Physical Kinetics} 
(Butterworth-Heinemann, Oxford, 1981).
\bibitem{r7} K. Kremer, M. O. Robbins, and G. S. Grest, Phys. Rev. Lett. {\bf 57}, 2694 (1986).
\bibitem{r8}R. O. Rosenberg and D. Thirumalai, Phys. Rev. A {\bf 33}, 4473 (1986).
\bibitem{r9} S. Hamaguchi and R. T. Farouki, J. Phys. {\bf 101}, 9876 (1994); S. Hamaguchi
{\it et al.}, Phys. Rev. E {\bf 56}, 4671 (1997).
\bibitem{r10} S. A. Khrapak {\it et al.}, Phys. Rev. Lett. {\bf 96}, 015001 (2006). 
\bibitem{r11} K. Avinash, Phys. Rev. Lett. {\bf 98}, 095003 (2007).
\bibitem{r12} P. Braun-Munzinger and J. Wambach, Rev. Mod. Phys. {\bf 81}, 1031 (2009).
\bibitem{r12a} B. Klumov, Phys. Usp. {\bf 53}, 1053 (2010).
\bibitem{r13} D. P. Resendes {\it et al.}, Phys. Lett. A {\bf 239}, 181 (1998).
\bibitem{r14} H. C. Lee {\it et al.}, Phys. Rev. E {\bf 56}, 4596 (1997).
\bibitem{r15} D. P. Resendes, Phys. Rev. E {\bf 61}, 793 (2000).
\bibitem{r16} S. L. Shapiro and S. L. Teukolsky, {\sl Black Holes, White Dwarfs and Neutron
Stars: The Physics of Compact Objects} (John Wiley \& Sons, New York, 1983). 
\bibitem{r17} H. M. van Horn, Science {\bf 252}, 384 (1991); V. E. Fortov, Phys. Usp. {\bf 52}, 615 (2009).
\bibitem{r17a} M. S. Murillo, Phys. Rev. E {\bf 81}, 036403 (2010); D. A. Chapman and D. O. Gericke,
Phys. Rev. Lett. {\bf 107}, 165004 (2011).
\bibitem{r17b} D. Kremp {\it et al.}, Phys. Rev. E {\bf 60}, 4725 (1999); 
V. M. Malkin, N. J. Fisch, and J. S. Wurtele, {\it ibid.} {\bf 75}, 026404 (2007);
H. Azechi {\it et al.}, Laser Part. Beams {\bf 9}, 193 (1991); R. Kodama {\it et al.}, 
Nature (London) {\bf 412}, 798 (2001).
\bibitem{r18} P. A. Markowich {\it et al.}, {\sl Semiconductor Equations} (Springer, Berlin, 1990).
\bibitem{r19}N. Crouseilles {\it et al.}, Phys. Rev. B {\bf 78}, 155412 (2008).
\bibitem{Gardner96} C. L. Gardner and C. Ringhofer, Phys. Rev. E {\bf 53}, 157 (1996).
\bibitem{Manfredi01} G. Manfredi and F. Haas, Phys. Rev. B {\bf 64}, 075316 (2001).
\bibitem{Tsintsadze09} N. L. Tsintsadze and L. N. Tsintsadze, EPL {\bf 88}, 35001 (2009).
\bibitem{r20} G. Manfredi, Fields Inst. Commun. {\bf 46}, 263 (2005).
\bibitem{r20b} J. T. Mendon\c{c}a, Phys. Plasmas {\bf 18}, 062101 (2011).
\bibitem{r21} P. K. Shukla and B. Eliasson, Rev. Mod. Phys. {\bf 83}, 885 (2011). 
\bibitem{r22} L. Brey {\it et al.}, Phys. Rev. B. {\bf 42}, 1240 (1990); See also
L. Hedin and B. I. Lundqvist, J. Phys. C: Solid State Phys. {\bf 4}, 2064 (1971).
It should be noted that the forces due to the electron exchange correlations 
are manifested in density functional theory (DFT) for an ensemble of electrons,
the wave functions of which follow a nonlocal Schr\"odinger equation.
An eikonal representation of the latter, similar to that in Ref. \cite{Tsintsadze09},
would yield Eq. (2) that includes the electrostatic, quantum, and electron-exchange/electron-correlation 
forces on equal footings.
\bibitem{r21a} S. V. Vladimirov and Yu. O. Tyshetskiy, Phys. Usp. {\bf 54}, 1243 (2011).
\bibitem{r21b} F. Haas, {\sl Quantum Plasmas: An Hydrodynamical Approach} (Springer, New York, 2011).
\bibitem{Else10} D. Else {\it et al.}, Phys. Rev. E {\bf 82}, 026410 (2010).
\bibitem{r23} P. K. Shukla and B. Eliasson, Phys. Lett. A {\bf 372}, 2897 (2008). 
\bibitem{Friedel58} J. Friedel, Adv. Phys. {\bf 3}, 446 (1954); Nuovo Cimento, Suppl. {\bf 7}, 287 (1958).
\bibitem{Kohn59} W. Kohn, Phys. Rev. Lett. {\bf 2}, 393 (1959).
\bibitem{Else11} D. Else {\it et al.}, EPL {\bf 94}, 35001 (2011).
\bibitem{Lindhard54} J. Lindhard, K. Dan. Vidensk. Selsk. Mat. Fys. Medd. {\bf 28}, 1 (1954).
\bibitem{r24} C. Kittel, {\sl Introduction to Solid State Physics} (John Wiley \& Sons, Inc. New York, 1986),  
p. 83.
\bibitem{Bonitz05} M. Bonitz {\it et al.}, Phys. Rev. Lett. {\bf 95}, 235006 (2005).
\bibitem{Bonitz08} M. Bonitz {\it et al.}, Phys. Plasmas {\bf 15}, 055704 (2008).
\end{thebibliography}
\end{document}